\documentclass[prb,preprint,amsmath,amssymb]{revtex4}

\usepackage[dvips]{color,graphicx}
\usepackage{graphicx,wrapfig}
\usepackage{amssymb}
\usepackage{bm}

\begin{document}

\title{Anomalous Hall conductivity and quantum friction}

\author{Pavel St\v{r}eda and Karel V\'{y}born\'{y}}

\affiliation{Institute of Physics, Academy of Sciences of the
Czech Republic, Cukrovarnick\'{a} 10, 16253 Praha 6,
Czech Republic
}

\date{Dec02, 2022}

\begin{abstract}
Anomalous Hall effect in high conductivity region
is studied using a two-dimensional network model. We find that the
off-diagonal conductivity comprises two parts: one which reflects the bulk
properties as obtained by the Kubo formula and another which is sensitive
to boundary conditions imposed on the network. In fully coherent
limit the latter scales with the width of the conducting channel,
while for real-world samples it is controlled by the coherence length.
It provides an alternative interpretation of the observed behavior in
the clean limit which is otherwise attributed to the skew scattering.
We highlight analogies to friction in viscous fluids responsible for
Couette flow. In the present case, this quantum effect is governed by
wave interference.
\end{abstract}


\maketitle

\section{Introduction}
Scattering is an essential ingredient to many transport phenomena.
The anomalous Hall conductivity $\sigma_{AH}$ of ferromagnetic systems provides
a notable exception to this rule but only in certain region: most materials
with {\em moderate} longitudinal conductivity $\sigma_{0}$ show almost constant
$\sigma_{AH}$ as scattering strength and hence $\sigma_0$ is varied.
This property seems to be well understood in terms of Berry curvature
of occupied electronic bands representing properties of ideal Bloch
systems , also called the intrinsic region of the 
anomalous Hall effect (AHE). This interval of roughly $10^4$ to $10^6$
inverse $\Omega.$cm is surrounded by regions in which $\sigma_{AH}(\sigma_0)$
becomes scattering dependent as reviewed by Nagaosa \textit{et al}.\cite{Nagaosa} 
While suppression of $\sigma_{AH}$ for stronger disorder is natural,
its linear increase with $\sigma_0$ in high conductivity
regions\cite{Miyasato} is,
at least, surprising. It is generally accepted that it is caused by
skew scattering,\cite{Smit,Luttinger,Onose,Sinitsin,Onoda}
asymmetric scattering of electrons on impurities induced by their
non-zero spin. Its effect increases with decreasing impurity
concentration. A seemingly inevitable consequence of this argumentation
is that anomalous Hall conductivity in clean systems is driven by
negligible impurity concentration while in this limit intrinsic
values obtained for ideal Bloch systems could be expected. With few
notable exceptions, the effect of Berry phases (see Eq.~\ref{Bloch_function}
below) is ignored in the context of skew scattering and 
even if it is not,\cite{Ishizuka:2017_a} the lack of generality
(see Appendix) leads to the contradiction mentioned above.
The main aim of the present treatment is to suggest another possible
origin of the observed increase of the Hall conductivity with
sample purity which does not rely on skew scattering.

Basic condition for the observation of AHE in magnetic systems
is the existence of non-zero orbital momentum\cite{note1}, which
can be induced by spin-orbit interaction or non-coplanar magnetic
order.\cite{Shindou:2001_a} It is responsible for the violation of time
reversal symmetry, a necessary condition for non-zero Hall effect.
Transport properties are measured on stripes, Hall bar samples, and
orbital momentum of atomic-type wave functions causes the
space current density oscillating across the stripe. It can
be represented by current paths with alternating current
directions.\cite{Jonckheere} Physically acceptable paths at
edges should be of the chiral type leading current along
opposite directions and the total net current thus vanishes
in the equilibrium. Voltage drop applied between stripe edges
induces changes of the electron concentrations within current
paths. It leads to the polarization of the system which is a
typical property accompanying the anomalous Hall effect.
\cite{Karplus,Adams,Fivaz} Coupling between current paths is
generally represented by their mutual friction. It defines
momentum transfer between edges as well as Hall current through
the stripe cross section. In quantum coherent systems such
friction is controlled by the wave interference. The main aim
of our approach to AHE is to show that quantum friction\cite{note2}
between chiral current paths can be responsible for a linear
increase of the anomalous Hall conductivity with $\sigma_0$
in the high conductivity region. It is an
extrinsic contribution due to the finite sample dimensions.

To verify this idea, a two-dimensional network model\cite{Streda_15}
will be used. It allows to apply theory of quantum
graphs\cite{graphs} ideally suited for studies of interference effects.
It contains all basic ingredients necessary for the existence of AHE.
Coupling between atomic orbitals is defined by $S$-matrix which is
convenient for application of the scattering matrix approach invented
by Landauer.\cite{Landauer57and70} Detailed model description and its
basic properties are presented in Section II.  The subsequent Section
is devoted to the properties of edge states. It will be shown that
chiral edge states crossing energy gaps which are responsible for the
quantum Hall effect\cite{niceREF-QHE} can be created or removed by
tuning  the boundary conditions. Contrary to the case of external
rational magnetic fields,\cite{Thouless} chirality is not determined
by wave function properties at the Brillouin zone boundaries.
The key part of our treatment is described in Section~IV where
scattering matrix approach is applied to obtain intrinsic Hall
conductivity and enhanced Hall current given by friction between
chiral current paths. In Section~V a brief summary of experimental
works on anomalous Hall conductivity in all three regions is given
and a two band model is used to obtain its qualitative features
for a large range of the system disorder strength. It is shown
that the experimentally observed behavior of $\sigma_{AH}(\sigma_0)$
can be reproduced without invoking the skew scattering mechanism.
The article will be completed by summary of main results and
concluding remarks.


\section{TWO-DIMENSIONAL NETWORK MODEL}
\begin{figure}[h]
\vspace{1mm}
\includegraphics[angle=0,width=3.3 in]{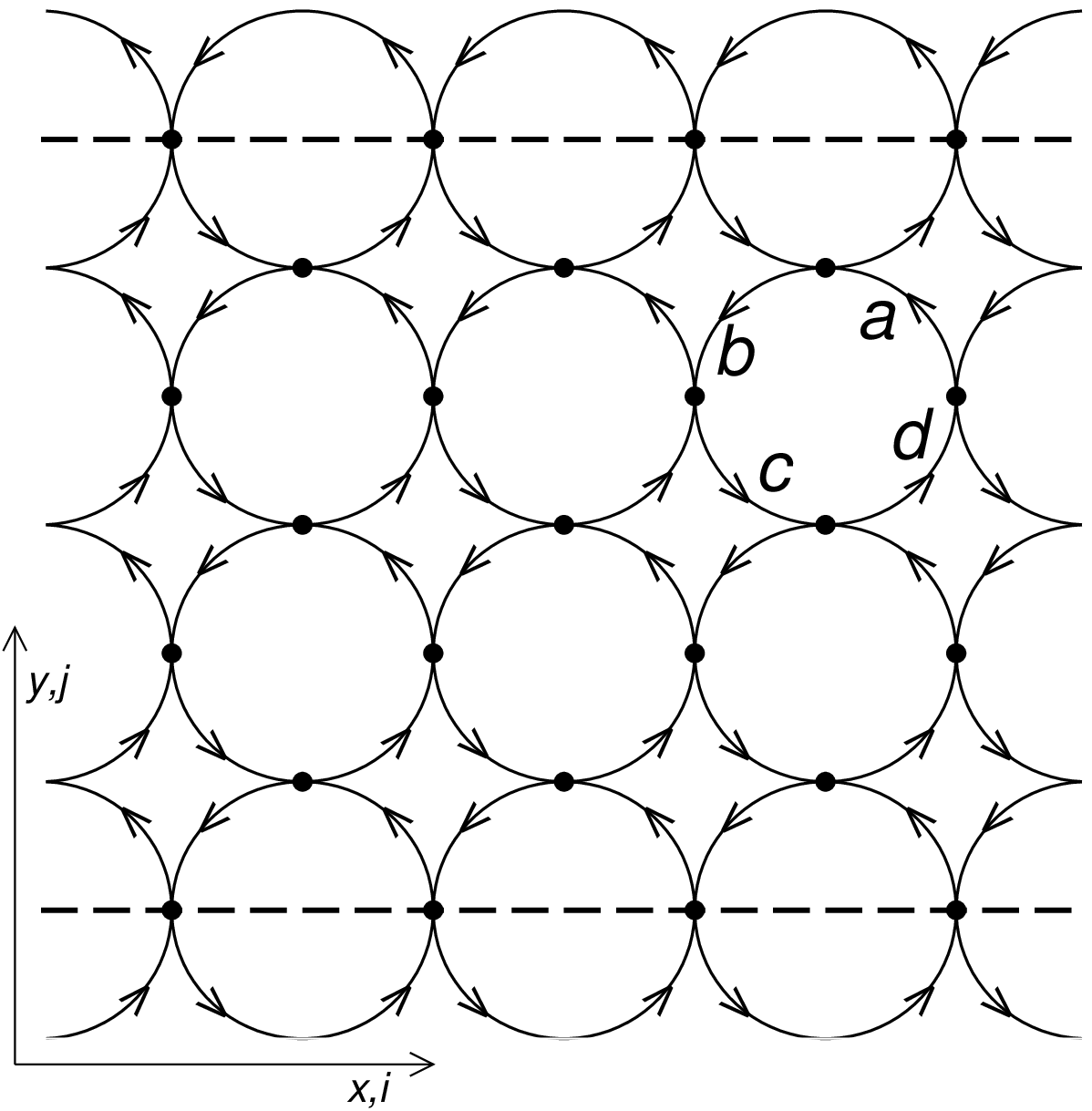}
\caption{Two-dimensional network model of coupled
orbitals with positive orbital momenta controlled by parameter
$\delta$. Arrows are indicating direction of the electron
motion and $\phi$-range of amplitudes $a,b,c,d$ are defined in
Tab.~\ref{tab-01}.}
\label{strip}
\end{figure}

In strictly two-dimensional systems spin-orbit interaction 
having form $L_zs_z$, $L_z$ being orbital momentum, separates
electrons into two independent groups having spin $s_z=1/2$ and
$s_z=-1/2$, respectively. Atomic state of the orbital number $m$
and spin $s_z$ has the same energy as that with $-m$ and $-s_z$.
Degeneracy of corresponding bands is removed by exchange interaction
which will be approximated by an effective Zeeman splitting. To
estimate general features of the anomalous conductivity the simple
two-dimensional network model sketched in Fig.~\ref{strip} has been
used \cite{Streda_15}. It allows to employ theory of graphs 
\cite{graphs} for single-mode quantum structure with $\delta$-type
coupling between orbitals. \cite{Exner} Such type of model graphs
(e.g. Chalker-Coddington model\cite{Chalker})
has already been applied to describe localization effect in
quantum Hall systems\cite{Marston:1999_a} and properties
of quantum spin-Hall systems. \cite{Kobayashi} 

Let us briefly recapitulate main ideas and basic properties of
the used model~\cite{Streda_15} on which our treatment is based.
Scattering matrix for individual contacts defines the
transmission probability $|t|^2$ representing the overlap integral
entering the standard tight-binding approach. The spin quantum number
allows to distinguish energy bands and define anomalous Hall
conductivity for each of the spin subsystems. For the sake of
simplicity, the spin parts of wave functions will not be shown
explicitly in the following treatment. We keep in mind that a
typical AHE setting will entail two copies of the network with
opposite spins and counter-propagating wavefunctions,
i.e. composed of orbitals with opposite angular momentum.

Atomic orbitals on individual lattice sites $\vec{R}_{i,j}$ are
modeled by rings of the radius $R$ formed by one-dimensional conductors.
Each electron subsystem (spin up and spin down) is represented by a one-way
conductor. Their eigenenergies and eigenfunctions
\begin{equation} \label{single_orbit}
E_m \, = \, \frac{\hbar^2 m^2}{2 m_0 R^2}
\quad , \quad
\psi_m(\phi) \, = \, \frac{1}{\sqrt{2 \pi R}} \; e^{i m \phi}
\; ,
\end{equation}
where $\phi\in(0,2\pi)$ is the polar angle are labeled by the
quantum number $m=0,\pm1, \ldots$ which defines angular momentum.
The assumption that electrons can orbit within rings
in one direction only leads to a non-zero orbital momentum,
and consequently removes the time reversal symmetry which is a
necessary condition for the Hall effects to emerge.

In the square lattice shown in Fig.~\ref{strip}, each of the rings has
four contact points with its neighbors which separate the domain of
the wave function amplitude  $A(\phi) \exp(i \delta \phi)$ into four 
sections listed in Tab.~\ref{tab-01}.  These allows to define four complex
amplitudes $a,b,c,d$ per lattice site fully determining the wave
function for given $\delta$.

\begin{table}
  \begin{tabular}{cc}
    amplitude & for $\phi\in$  \\ \hline
    $a_{i,j}$ & $(0,\pi/2)$ \\
    $b_{i,j}$ & $(\pi/2,\pi)$\\
    $c_{i,j}$ & $(\pi,3\pi/2)$\\
    $d_{i,j}$ & $(3\pi/2,2\pi)$
  \end{tabular}
  \caption{Definition of $a,b,c,d$ for piece-wise constant $A(\phi)$
  pertaining to the ring at $\vec{R}_{i,j}$.}
  \label{tab-01}
\end{table}

Considering the positive orbital momenta of the
atomic-type orbitals,  $\delta > 0$, the amplitudes are controlled by
the following relations:
\begin{eqnarray} \label{basic_Eq_set} 
\begin{array}{ccc}
e^{-i \delta \pi} \, a_{i,j} &=& r \, e^{i \delta \pi} \, d_{i,j} +
t \, b_{i+1,j}
\; ,
\\
b_{i,j} &=& r \, a_{i,j} + t \, e^{i \delta \pi} \, c_{i,j+1}
\; ,
\\
c_{i,j} &=& r \, b_{i,j} + t \, e^{i \delta \pi} \, d_{i-1,j}
\; ,
\\
d_{i,j} &=& r \, c_{i,j} + t \, e^{-i \delta \pi} \, a_{i,j-1}
\; ,
\end{array}
\end{eqnarray}
where $t$ denotes transition coefficient of the wave entering adjacent
orbital while $r$ represents part of the  wave continuing the orbital
motion. For the considered $\delta$-type coupling they are of the following
general form \cite{Exner}
\begin{equation}
t = \frac{i \alpha}{1- i \alpha}
\quad , \quad
r = \frac{1}{1- i \alpha}
\quad , \quad
|r|^2+|t|^2 =1
\; ,
\end{equation}
where $\alpha$ is a real parameter which is supposed to be an
energy independent constant for the sake of the simplicity.

For infinite periodic network the wave functions are of the Bloch form
\begin{eqnarray} \label{Bloch_function}
|m,\vec{k} \rangle \equiv \Psi_{m,\vec{k}}(\vec{r}) \, = \,
\frac{e^{i \theta_{m}(\vec{k})}}{\sqrt{N}}
\sum_{i,j=1}^N e^{i \vec{k} \vec{R}_{i,j}}
\sqrt{\frac{1}{2 \pi R}} \, \times 
\nonumber \\ 
\times
A_{m,\vec{k}}(\phi) \, e^{i \delta_{\vec{k}} \phi}
\delta(|\vec{r}-\vec{R}_{i,j}| - R)
\equiv
e^{i \vec{k} \vec{r}} \, u_{m,\vec{k}}(\vec{r})
\; ,
\end{eqnarray}
where $m$ and $\vec{k}$ are the band number and the wave vector,
respectively, $\theta_{m}(\vec{k})$ denotes the Berry phase\cite{Xiao:2010_a}
and $u_{m,\vec{k}}(\vec{r})$ stands for the periodic part of Bloch
functions.
Wave function amplitudes $A_{m,\vec{k}}(\phi)$ are
subject to the following Bloch conditions
\begin{eqnarray}
\begin{array}{ccc}
c_{i,j+1} = e^{ik_y} \, c_{i,j}
\quad &,& \quad
a_{i,j-1} = e^{-ik_y} \, a_{i,j}
\; ,
\\
b_{i+1,j} = e^{ik_x} \, b_{i,j}
\quad &,& \quad
d_{i-1,j} = e^{-ik_x} \, d_{i,j}
\; 
\end{array}
\end{eqnarray}
where the wave vector components $k_{x,y}$ range from $-\pi$ to $\pi$
pursuant to the choice of units, lattice constant $a_0=1$.
Zero determinant of the resulting equations for wave function amplitudes
yields the spectral condition for dimensionless parameter $\delta_{\vec{k}}$
\begin{equation} \label{spect_cond}
\cos k_x + \cos k_y = -2 \, \cos{\delta_{\vec{k}} \pi} -
\frac{1-\alpha^2}{\alpha} \; \sin{\delta_{\vec{k}} \pi}
\; ,
\end{equation}
which can be transformed into dispersion relation for eigenenergies using
\begin{equation} \label{cr_dispersion}
E_{\vec{k}} \, = \, \frac{2 \hbar^2 \delta_{\vec{k}}^2}{m_0}
\end{equation}
in analogy to~(\ref{single_orbit}). This implicit expression for
energy $E_{\vec{k}}$ corresponds to the kinetic energy of
wave function~(\ref{Bloch_function}).

For any energy-independent value of the parameter $\alpha$, the spectrum
comprises a series of non-overlapping bands linked to states having orbital
number $m$.
Dispersions of the dimensionless
parameter $\delta \sim \sqrt{E}$ are periodic with the period 2. Change of
$\delta$ by one gives the same dispersion but shifted in $\vec{k}$-space
by $\Delta \vec{k} = (\pi , \pi)$. All these features can be seen in 
spectrum obtained for a stripe samples shown in Fig.~\ref{edgestates}.
Energy gaps become closed for $|t|^2$ approaching the value $|t|^2 = 0.5$
($\alpha= \pm 1$) at which the band width equals to that defined by
$\Delta \delta =1$. Fixed-energy contours (see Fig.~2 in
Ref.~\onlinecite{Streda_15}) are identical to the case of cosine
band dispersion produced by the well-known square-lattice tight-binding
model but energy scaling~(\ref{cr_dispersion}) is different.

In most real-world systems, the coupling between adjacent atomic
orbitals is weaker than that to the atomic core and crystal formation
lowers energy of electron states. For these reasons $|t|^2 < 0.5$
and  $\alpha > 0$ satisfying the above requirements will be
preferred in the following treatment. For $|t|^2 > 0.5$,
electrons 'orbit the stars' rather than the circular orbitals around
$\vec{R}_{i,j}$, i.e. they are pushed into the interstitial positions.
In these cases the coupling gives rise to energy of electron states.
For $|t|^2=1-|r|^2=0,1$ the limit of isolated orbitals is achieved
whereupon the dispersions reduce to flat bands.

\section{EDGE STATES} \label{SecEdgestates}

Quantization of the anomalous Hall conductivity has also been
observed on systems endowed with non-zero orbital
momentum.\cite{Berne,Konig,Chang:2013_a}
Generally it is attributed to the existence of chiral edge states
within gap regions, i.e. states having opposite velocity at opposite
sample edges. Existence of such states has been first predicted for
two dimensional systems subjected to a strong external magnetic
field. In this case there are two scaling areas, the area per unit
magnetic flux $A_{\phi}$ and the unit cell area $A_0$.
For rational values of $A_0/A_{\phi}$ eigenfuctions are of the
Bloch form but the corresponding translation symmetry differs
from that at zero magnetic field. As it has been shown by
Thouless \textit{et al}. \cite{Thouless} number of chiral edge states,
Chern number, is fully determined by eigenfunction properties
at the Brillouin zone boundary. External magnetic field induces
orbital momentum of atomic type states leading to an increase of
the system energy. Chiral edge states are induced to minimize
it. For this reason they are insensitive to the boundary
conditions at the sample edges. \cite{Streda_1994} These general
arguments are not applicable in the zero field limit. Using two
dimensional network model the decisive role of boundary
conditions for existence of chiral edge states will be shown.

A stripe open along the $\hat{x}$ direction parallel to
main crystallographic axis will be considered. Bloch conditions in
the $\hat{x}$ direction, $b_{i+1,j}=e^{ik_x} \, b_{i,j}$ and
$d_{i-1,j}=e^{-ik_x} \, d_{i,j}$,
inserted into the basic equation set (\ref{basic_Eq_set}) give
\begin{eqnarray} \label{stripEqs}
\begin{array}{ccc}
- e^{-i \delta \pi} \, a_{i,j} + t \, e^{ik_x} \, b_{i,j} +
r \, e^{i \delta \pi} \, d_{i,j} &=& 0
\; ,
\\
r \, e^{-i \delta \pi} \,  a_{i,j} - e^{-i \delta \pi} \, b_{i,j} +
t \, c_{i,j+1} &=& 0
\; ,
\\
r \, e^{-i \delta \pi} \,  b_{i,j} - e^{-i \delta \pi} \, c_{i,j} +
t \, e^{-ik_x} \, d_{i,j} &=& 0
\; ,
\\
t \, a_{i,j-1}  + r \, e^{i \delta \pi} \,  c_{i,j} -
e^{i \delta \pi} \, d_{i,j} &=& 0
\; .
\end{array}
\end{eqnarray}
For a given $k_x$ the eigenvalue problem reduces to the problem for
a single column of orbitals. It is independent of its position defined
by the index $i$. Two types of boundary conditions in the $\hat{y}$
direction will be considered: (i) hard walls leaving circular orbitals
untouched and (ii) those which cut orbitals in half as shown in
Fig.~\ref{strip} by dashed lines. Electrons are thus skimming or skipping
along stripe walls.

\begin{figure}[h]
\vspace{1mm}
\includegraphics[angle=0,width=3.3 in]{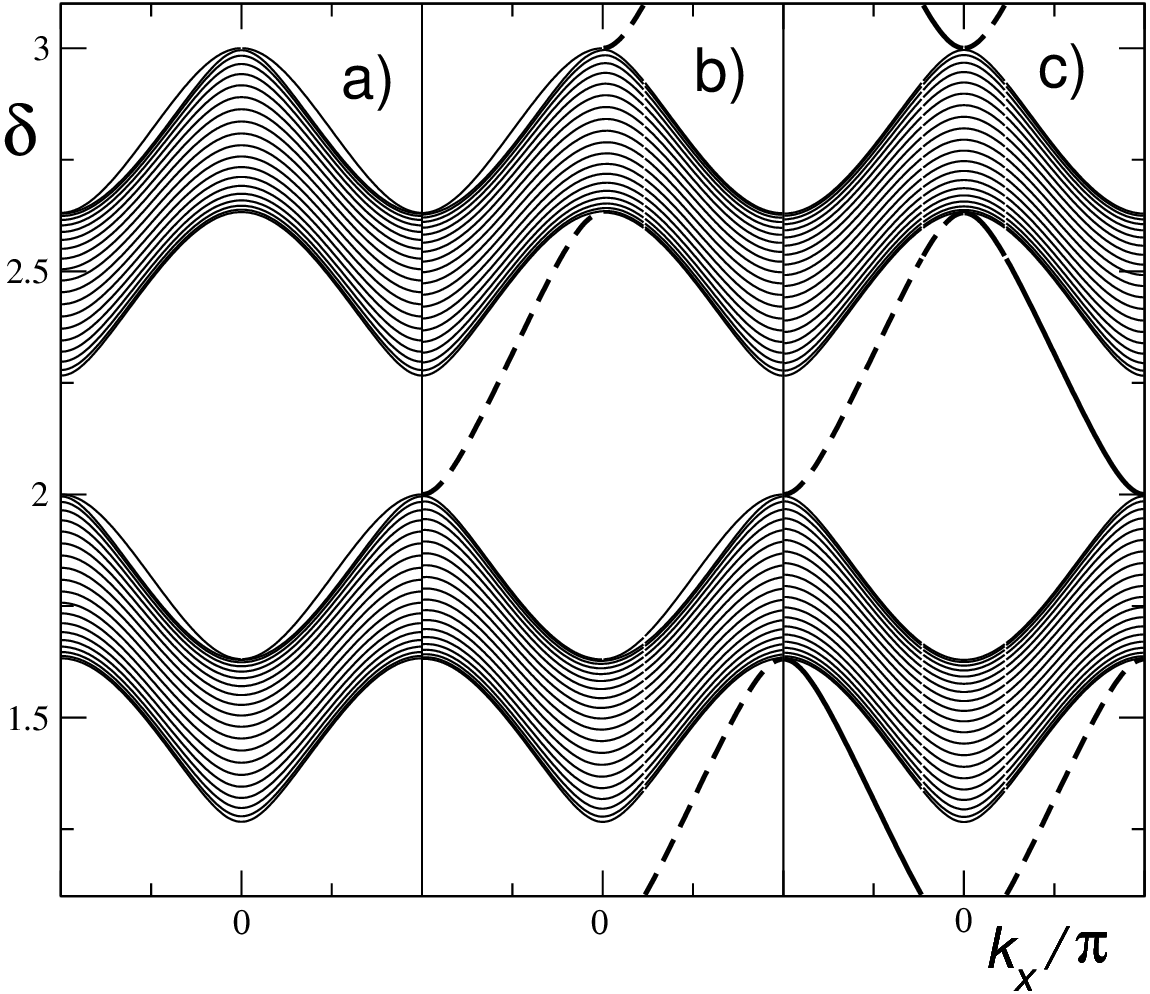}
\caption{Energy branches for a stripe subjected to different boundary
conditions, $N=20$, $|t|^2=0.3$ and $\alpha > 0$. Thick full and dashed
lines represent edge states at lower and upper stripe edges, respectively.}
\bigskip
\label{edgestates}
\end{figure}

Branch dispersions representing the case (i) for the column composed of
$N=20$ circular orbitals controlled by the boundary conditions
$b_{i,N}=a_{i,N}$ and $d_{i,1}= c_{i,1}$ are shown in
Fig.~\ref{edgestates}a. At any band energy the electron path at the
upper edge ($\cdots \rightarrow a_{i,N} \rightarrow b_{i,N}
\rightarrow a_{i-1,N} \rightarrow \cdots$) and that at the lower
edge ($\cdots \rightarrow c_{i,1} \rightarrow d_{i,1} \rightarrow
c_{i+1,1} \rightarrow \cdots$) carry skimming electrons in opposite
directions, see Fig.~\ref{strip}.

As for the case (ii) chiral edge states crossing the energy gaps
appear. For stripe of the width $N a_0$ the boundary conditions
$e^{i \delta \pi} d_{i,N+1}=c_{i+1,N+1}= e^{i k_x} \, c_{i,N+1}$ and
$e^{i \delta \pi} b_{i,1}=a_{i-1,1}=e^{-ik_x} \, a_{i,1}$ correspond
to hard walls cutting the the orbitals in half on both sides of the
sample. Resulting branch dispersions are shown in Fig.~\ref{edgestates}c
for $N=20$. In real space, electrons at edge current paths are
skipping along stripe walls flowing in opposite directions compared
to the previous case of skimming electrons.

There exists a peculiar possibility of imposing mixed boundary
conditions, type (i)/(ii) at the lower/upper edge, giving rise to
energy dispersions shown in  Fig.~\ref{edgestates}b. In this case
the symmetry leading to the presence of chiral edge states is lost
and for chemical potential within the gap region the current flow
is allowed only along one edge. It represents an ideal diode.
Comparison of all three cases suggests that edge states are
exclusively determined by boundary conditions. This conclusion is
supported by a close connection of the anomalous Hall effect to the
polarization which is known to be affected by boundary conditions.
Opposite to the case of external magnetic fields (corresponding to
rational values of $A_0/A_\phi$), the appearance of chiral edge
states is not determined by the Chern number which in our case has
zero value. Necessary conditions for their appearance are the chiral
symmetry of the current distribution across the stripe, i.e.
oscillating currents are surrounded by current paths at the stripe
edges leading currents in opposite direction, and relevant boundary
conditions.

Note that for transition probability $|t|^2 > 0.5$
edge states crossing energy gaps appear only at edges for which
hard wall leaves circular orbitals untouched.
Nevertheless general conclusions remain unchanged.


\section{ELECTRONIC TRANSPORT: SCATTERING MATRIX APPROACH}
Corbino disc samples can be used to measure conductivity components
directly, at least in principle. In the limit of the infinite disc
radius it is equivalent to a stripe open in one direction (in our case
$\hat{x}$) coinciding with a main crystallographic axis. A voltage drop
applied to the opposite stripe edges induces current which has two
components, perpendicular and parallel with the edges representing
longitudinal (i.e. along $\hat{y}$) and Hall currents, respectively.
Scattering matrix approach \cite{Landauer57and70,LB} will be used to 
evaluate corresponding conductivities, $\sigma_{yy}$ and 
$\sigma_{xy}$, for the already described two-dimensional
network model. It represents response to the electron concentration
gradient of the fully coherent system, i.e. no dissipation is
allowed within stripe interior. Dissipation is supposed to take place
at the source and drain only where electrons are subjected to the
equilibration processes.  Stripe width can thus be identified with
the equilibration length $\lambda_e$. Electron wave functions are
thereby also losing information about their phases and the coherence
length $\lambda_c$ thus coincides with the stripe width as well.

\begin{figure}[h]
\vspace{1mm}
\includegraphics[angle=0,width=3.3 in]{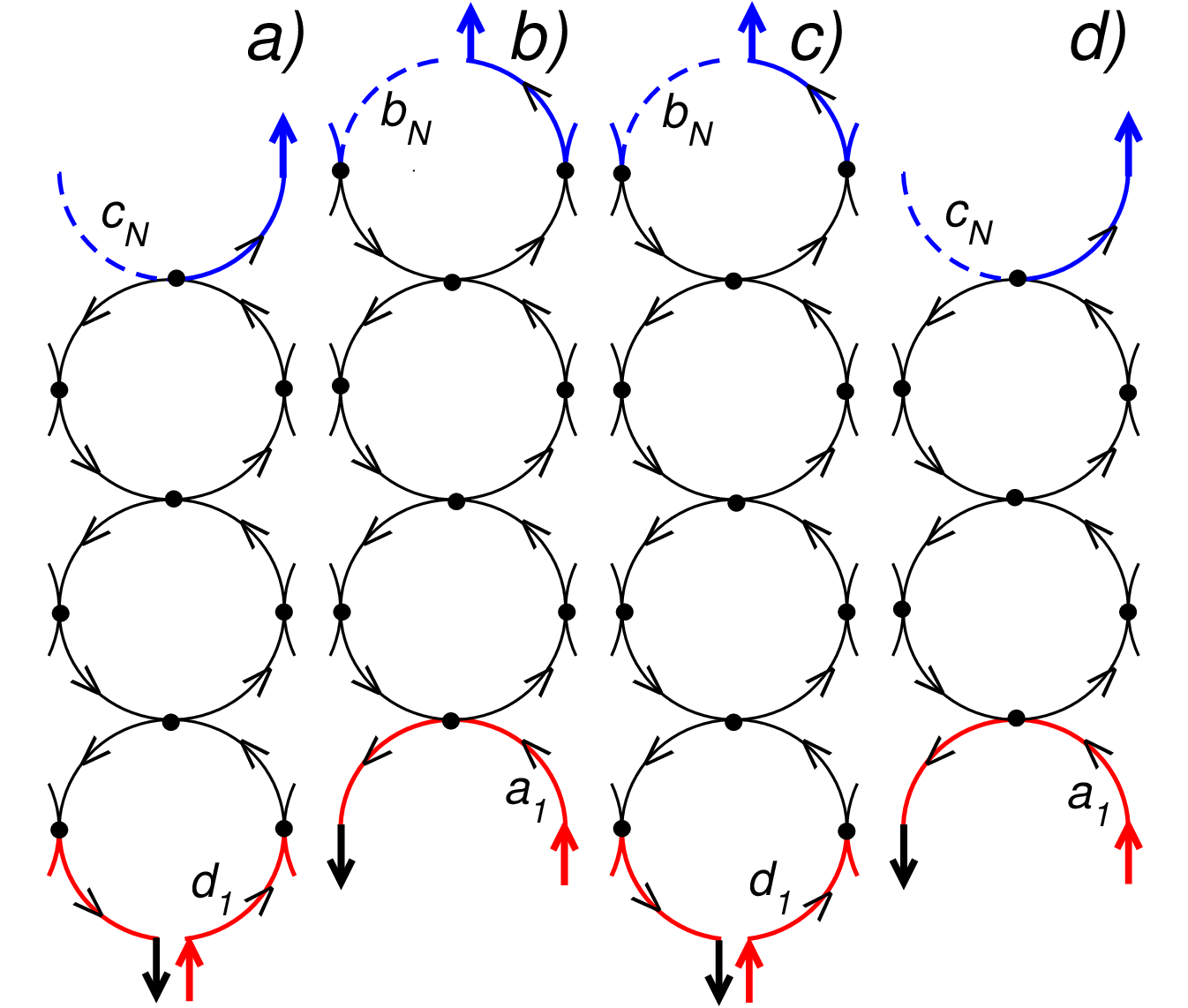}
\caption{(color on line) Scheme of four possible boundary conditions
applied to a single column. Source and drain are marked in red and blue,
respectively.}
\label{columns}
\end{figure}

Stripe interior is composed of electron paths leading currents along
positive or negative $\hat{x}$ direction. To analyze conductivity
contributions within the stripe, it is natural to choose one of the
electron paths as the source and another as the drain, shown in Figures
by red and blue lines, respectively. Among the four possibilities sketched
in Fig.~\ref{columns} there are two qualitatively distinct cases.
Source and drain paths can be chosen to carry current along the
same direction (we choose to call it the Born-von K\'{a}rm\'{a}n case)
or their currents have opposite direction (the chiral case).
These two cases will be treated separately in following Subsections and
the Hall conductivity of very clean but not fully coherent systems will
be discussed in the last Subsection. 

Bloch conditions along $\hat{x}$ direction reduce problem to scattering
within the single column of orbitals for each of the wave numbers $k_x$.
Wave function amplitudes are defined by Eq.~(\ref{stripEqs}) accompanied
by appropriate current currying conditions. Averaging over $k_x$ gives
relevant results. To get smooth enough dependence on the parameter $\delta$
defining the energy a large number of $k_x$ values has to be used.
Usually $10^4-10^5$ $k_x$-values uniformly spread through
the interval $k_x \, \epsilon \, (-\pi,\pi)$ are considered. Results
of the scattering matrix approach do not depend on the column position
(i.e. index $i$) and unless necessary, this index will be skipped for
brevity. For presented numerical examples, unless explicitly stated,
the model parameters $\alpha>0$ and $|t|^2=0.3$ will be considered.

\subsection{Born-von K\'{a}rm\'{a}n cases}
\begin{figure}[h]
\vspace{1mm}
\includegraphics[angle=0,width=3.3 in]{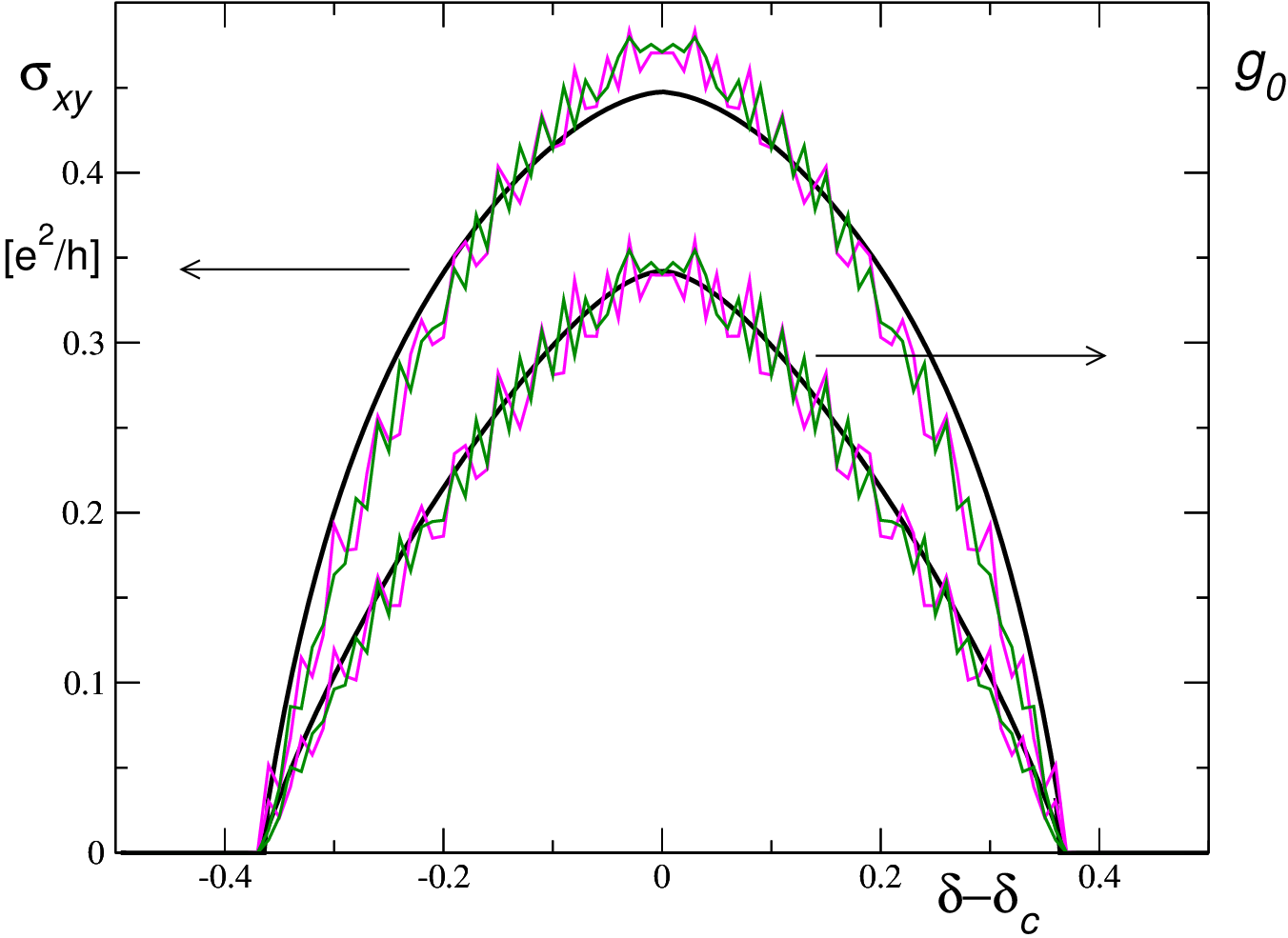}
\caption{(color on line) Conductance $g_0(\delta)$ and anomalous Hall
conductivity $\sigma_{xy}(\delta)$ obtained by using scattering matrix
approach in the Born-von K\'{a}rm\'{a}n case.
Magenta and green curves correspond to the column lengths
$N=15, \, 25$ and $|t|^2=0.3$. Smooth thick
lines are Kubo formula results for unbounded systems.}
\label{g0H_BK}
\end{figure}

Let us first consider source and drain paths at which electron
velocity along $\hat{x}$ direction is positive, as sketched in
Fig.~\ref{columns} a). Electrons are supposed to be injected
into strip region via the lower path, $d_{1}=1$, while they
are absorbed by upper path and the condition $c_{N}=0$ ensures
zero injection from this side. Equation set (\ref{stripEqs})
together with these  conditions define uniquely all
amplitudes within the column. Transition coefficient for given
$k_x$ is given by the amplitude $d_{N}(k_x,\delta)$. Total
transition probability $T(\delta)$ defining current flow through
a single column along $\hat{y}$ direction reads
\begin{equation} \label{SM_g0}
T(\delta) \equiv
\frac{h}{e^2} \, g_0(\delta) = \left\langle |d_{N}(k_x,\delta)|^2
\right\rangle_{k_x}
\; ,
\end{equation}
where $g_0(\delta)$ stands for conductance per single column.
In this case transitions between orbitals are independent on their
position. Except for fluctuations
due to the size quantization it is independent of the considered
column length $(N-1/2)a_0$ as the green and magenta
curves in Fig.~\ref{g0H_BK} show.
Considering an energy independent relaxation time represented by
the parameter $\gamma$, the Kubo formula for longitudinal conductivity
$\sigma_0(\delta)$ given by Eq.~(\ref{sig0_Kubo}) can formally be
fit to approximate $g_0(\delta)$, as shown in Fig~\ref{g0H_BK}.

Current flow along $\hat{x}$ direction representing Hall current
is not uniformly spread through the sample  cross-section indicating
the decisive role of wave function phases.
Hall conductivity can be defined as follows
\begin{equation} \label{BK_gH}
\sigma_{AH}\equiv
\sigma_{xy}(\delta) = \! \frac{e^2}{h} \sum_{j=1}^{N} \!
\left\langle |d_{j}(k_x,\delta)|^2 \! - |a_{j}(k_x,\delta)|^2
\right\rangle_{k_x} \! \approx
\sigma_{xy}^{\rm (int)}(\delta)
\; ,
\end{equation}
where the sum over $j$ defines current into the right hand side
column through coupling points enhanced by the current within drain
path defined by setting $a_N(k_x,\delta)=0$. Drain contribution decreases
with rising column length and currents through coupling points
becomes dominant. Again, except for the size quantization effect the
obtained results are independent on the column length and they are
close to the intrinsic Hall conductivity 
$\sigma_{xy}^{\rm (int)}(\delta)$ given by Eq.~(\ref{Bloch_Bastin}),
as shown in Fig.~\ref{g0H_BK}. This comparison entails no fitting
procedure. Note that conductance $g_0(\delta)$ defines the current
through the unit cell cross-section and corresponding density is thus
much larger than the Hall current density.

Another possibility is to choose source and drain paths leading electrons
along negative direction as shown in Fig.~\ref{columns} b). In this
case ($a_1=1$ and $b_N=0$) the conductance $g_0(\delta)$ defined by
probabilities $|a_N(k_x,\delta)|^2$ coincides with that determined in
the previous case.
To obtain Hall conductivity instead of the drain path contribution
the current of the source path has to be added by setting
$d_1(k_x,\delta)=0$. In this case equation (\ref{BK_gH}) gives for
energy gap regions quantum value $-e^2/h$. This edge state effect
has to be subtracted to obtain Hall conductivity within the bulk.
Resulting $\delta$-dependence then coincides with that of the
previous case.

\subsection{Chiral cases}
Transport between source and drain electron paths which carry current
in opposite directions leads to qualitatively different results.
Considering columns of length $N a_0$, the relevant
conditions are $d_1=1$ and $b_N=0$ or $a_1=1$ and $c_{N}=0$,
as sketched in Fig.~\ref{columns} c) and d), respectively.

Except for fluctuations caused by size quantization, the transition
probability in both cases is once more independent of the column
length. Compared to the Born-von K\'{a}rm\'{a}n case, conductance
$g_0$ per single column is $\beta$-times smaller. This ratio
is only weakly $|t|$-dependent, for example for
$|t|^2=0.3$ and $|t|^2=0.2$ the ratio $\beta$ equals to $0.83$ and
$0.87$, respectively.

\begin{figure}[h]
\vspace{1mm}
\includegraphics[angle=0,width=3.3 in]{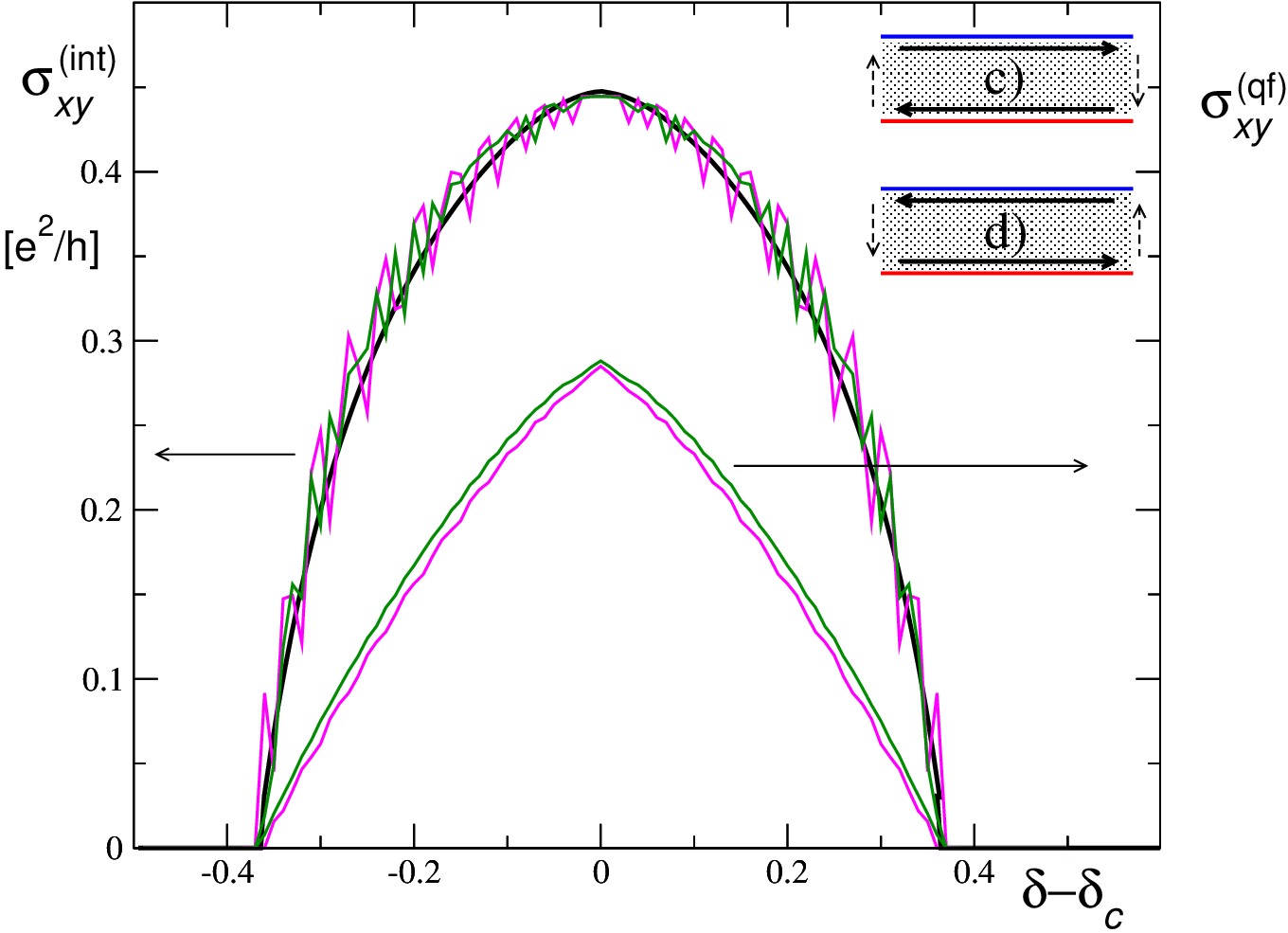}
\caption{(color on line) Anomalous Hall conductivity contributions
$\langle{\sigma}^{\rm (Ch)}_{xy}(\delta) \rangle/\beta$ and
$\sigma_{xy}^{\rm (qf)}(\delta)/\beta$ obtained by using
scattering matrix approach with chiral type boundary
conditions. The two chiral scenarios are sketched in the inset whereas
the letters refer to situations shown in Fig.\ref{columns}.
Magenta and green curves correspond to $N=15, \, 25$,
scaling factor $\beta=0.83$ is discussed in the text and $|t|^2=0.3$.
Smooth thick line is the Kubo formula result. }
\label{gH_CH}
\end{figure}

Essential difference from the Born-von K\'{a}rm\'{a}n case is the
dependence of the Hall conductivity on the distance between source and
drain $N_s a_0$ which reads
\begin{eqnarray} \label{CH_enhancement}
\sigma_{xy}^{\rm (\pm Ch)}(\delta) &=&
\langle \sigma_{xy}^{\rm (Ch)}(\delta) \rangle
\pm N_s \sigma_{xy}^{\rm (qf)}(\delta) \equiv
\nonumber \\
&\equiv&
\langle \sigma_{xy}^{\rm (Ch)} (\delta) \rangle \pm
\Delta \sigma_{xy}^{\rm (Ch)}(\delta)
\; ,
\end{eqnarray}
where
\begin{equation}
\langle \sigma_{xy}^{\rm (Ch)}(\delta) \rangle \approx \beta
\, \sigma_{xy}^{\rm(int)} (\delta)
\; ,
\end{equation}
denotes the average value of both chiral cases, c) and d).
Plus and minus sign correspond to scattering
problems with opposite chirality of current paths as sketched in
the inset of Fig.~\ref{gH_CH}.
It is determined by velocity sign of electrons within
the path attached to the source path. Contribution per unit
cell $\sigma_{xy}^{\rm (qf)}(\delta)$ represents average friction
between the nearest current paths. For large enough $N_s$ it reaches
a constant value as illustrated in Fig.~\ref{gH_CH}.

The ratio $\beta$ for the intrinsic part
$\langle \sigma_{xy}^{\rm (Ch)}(\delta) \rangle$  turns out to be
the same as for the longitudinal conductance. Its deviation of from
one is the result of wave interference modified by the change of
the boundary conditions.

Hall current enhancement $\Delta \sigma_{xy}^{\rm (Ch)}(\delta)$
entering Eq.~(\ref{CH_enhancement}) can be
understood using the analogy with the viscous flow in classical fluids.
The largest current is flowing via the path attached to source.
It is stirring currents within adjacent stripe paths forcing them
to move along the same direction. It explains the origin of the
enhanced Hall current and in particular, its direction. 
In the fully coherent systems the
friction between current paths is determined not only by average
coupling $|t|^2$ but it is modified by the wave interference which
determines coupling between distant current paths.

\subsection{Anomalous Hall effect in high conductivity region}

Let us first discuss Hall conductivity within a coherent area
of infinite systems. It has been found that the conductance $g_0(\delta)$
between paths leading current in opposite directions (chiral cases
c) and d) in Fig.~\ref{columns}) is smaller than that for which they are
flowing along the same direction (Born--von K\'{a}rm\'{a}n cases a) and b)
in the same figure). The occupation of neighboring paths
leading current in opposite directions, has to be appropriately
modified to unify both current densities. This polarization accompanying
anomalous Hall effect allows electron transfer without enforced
dissipation due to the current differences. It also ensures that
intrinsic parts of the Hall conductivity is the same in both cases.
Resulting Hall conductivity is given by average value of all
four contributions discussed in previous two Subsections. They
are of the same probability to appear and consequently sum of
friction contributions depending on the column length is averaged
out. Resulting anomalous Hall conductivity approaches intrinsic
values given by the Kubo formula.

Different results are obtained for stripe samples shown in
Fig.~\ref{strip} with two types of boundary conditions discussed
in the previous Section~III as (i) and (ii) cases. They are
composed of columns containing integer number of unit cells.
In these cases the edge electron paths carry current in opposite
directions. Let us assume that attached source and drain prepared
from the same material are coherently coupled to the stripe electron
system which corresponds to scattering problems shown in
Fig.~\ref{columns} c) and d). In the fully coherent case the Hall
current enhancement defined by Eq.~(\ref{CH_enhancement}) is
proportional to the stripe width $N_sa_0$ which for large enough
$N_s$ dominates.

Because of equilibration processes within source and drain, electrons
are losing all information about their past. The stripe width $N_sa_0$
can be thus identified with the equilibration length $\lambda_e$.
Longitudinal conductivity $\sigma_0 \equiv \sigma_{yy}$ can
be approximated by the conductance per square area $N_s^2 a_0^2$,
$\sigma_0=N_s g_0$, and the Hall conductivity enhancement
defined by Eq.~(\ref{CH_enhancement}) increases with $\sigma_0$ as
observed in the high conductivity region.

Classical analogy of this effect is Couette flow observed in fluids
placed between two plates. Motion of one plate induces fluid flow
along the same direction which in the stationary case decreases
linearly towards fixed one. In our case the role of the moving
plate is played by the current path attached to the electron source.

Dissipation processes are minimizing deviation from the
equilibrium. They are thus trying to suppress enhanced Hall current
by electron transitions into paths leading current in opposite
direction. For strong enough dissipation it can be thus expected
that quantum friction contributions will be averaged out giving
rise to intrinsic values of the Hall conductivity. Since current
enhancement originates in wave interference even low angle inelastic
scattering can be quite effective. It can be expected that
corresponding relaxation time $\tau_{qf}$ could be much smaller
than $\tau_e \propto \lambda_e$ which controls the longitudinal
conductivity. Contrary to the case of unbounded systems the effect
of transitions between chiral paths giving rise to opposite directions
of the Hall current enhancements cannot be averaged out since for
considered stripes their numbers differ by one.
For corresponding current contribution, Eq.~(\ref{CH_enhancement})
can be used with $a_0 N_s$ replaced by the coherence length
$\lambda_{c} \propto \tau_{qf}$. Anomalous Hall conductivity measured
on a stripe of width $w$ is thus given by averaged current density and we get
\begin{equation} \label{fric}
\sigma_{xy}^{\rm (\pm Ch)}(\delta) \approx  \sigma_{xy}^{\rm (int)}(\delta)
\pm \sigma_{xy}^{\rm (qf)}(\delta) \, \frac{\lambda_{c}}{w}
\; ,
\end{equation}
where plus and minus sign correspond to boundary conditions for which
electrons are skimming or skipping along strip edges, respectively.
Note that for $|t|^2 > 0.5$ the Hall conductivity has opposite sign
but its general features remain unchanged.

Estimation of the measured Hall conductivity given by Eq.~(\ref{fric})
has to be viewed as a rough approximation based on the assumption that
the enhanced current distribution is spread uniformly through width $w$.
If it becomes concentrated within a
slab of the width $w_{qf}$ at the edge vicinity the measured $\sigma_{xy}$
becomes affected by the ratio $w_{qf}/w$. This problem desires a more
advanced theoretical description based for example on the application
of non-equilibrium Green's functions\cite{Kovalev:2008_a} employed in
finite size systems. 

Note that analyzed Hall currents are spin polarized. For negative
values of $\delta$ the orbital momentum and the Hall conductivity
change their sign. Consequently spin polarization of Hall currents
is changed as well.

\section{Two band model}
As already mentioned in the introduction three regions of the anomalous
Hall conductivity $\sigma_{AH}$ in dependence on the disorder strength
represented by the longitudinal conductivity $\sigma_0$ can be
identified.\cite{Nagaosa} Scaling $\sigma_{AH}\propto \sigma_0^\nu$ in
the dirty-metal region with $\nu \approx \, 1.6$ has received considerable
attention for $\sigma_0$ down below units of inverse
$\Omega \cdot$cm.~\cite{Miyasato,FP08,Sangiao:2009_a} Phonon assisted
hopping between impurity localized states \cite{Sinova} gives
the observed scaling. Empirically,
there appears a transition from $\sigma_{AH}\propto \sigma_0^\nu$ to
the intrinsic region, $\sigma_{AH}\sim$~const., for $\sigma_0$ between
$10^3$ and $10^4$ inverse $\Omega \cdot$cm.\cite{Onoda}
It is attributed to suppression of the band overlaps with decreasing
disorder strength. \cite{Streda_2010,Naito}   Calculations
of intrinsic $\sigma_{AH}$ values for intermediate $\sigma_0$
(often in terms of Berry curvature) are a popular topic for ab initio
studies of ideal crystal structures\cite{Jungwirth:2002_a,Yao:2004_a,
Fuh:2011_a,Tung:2012_a,DK13,Chen:2014_a,Nayak:2016_a,Manna:2018_a,Helman:2021_a}
and even alloys\cite{Zhang:2013_a,Turek:2012_a} have been considered.
While there is abundance of experimental data for systems falling into
these two categories, data for high conductivity regions for
which conductivity is well above $10^5$ ($\Omega \cdot$cm)$^{-1}$ are
scarce.\cite{Miyasato,Majumdar:1973_a,Schad:1998_a,Shiomi,Xu,Omori:2019_a}
They require crystal structures with minimum lattice imperfections and
low temperatures to suppress dynamical disorder due to the electron
scattering with phonons and magnons.
Outstanding bulk samples of iron\cite{Majumdar:1973_a} with $\sigma_0$ in
excess of $10^8$ ($\Omega\cdot$cm)$^{-1}$ showed an increase of
$\sigma_{AH}$ with $\sigma_0$, and the same was observed\cite{Schad:1998_a} 
for thin layers of somewhat lower quality. Newer study\cite{Miyasato}
confirms this and reports a decrease of $\sigma_{AH}$ for cobalt rather
than the increase seen in iron. This work shows almost constant
$\sigma_{AH}$ for nickel down to the lowest $\sigma_0$ achieved but
better-conductivity samples\cite{Xu} still show some increase in
$\sigma_{AH}$.

\begin{figure}[h]
\vspace{1mm}
\includegraphics[angle=0,width=3.3 in]{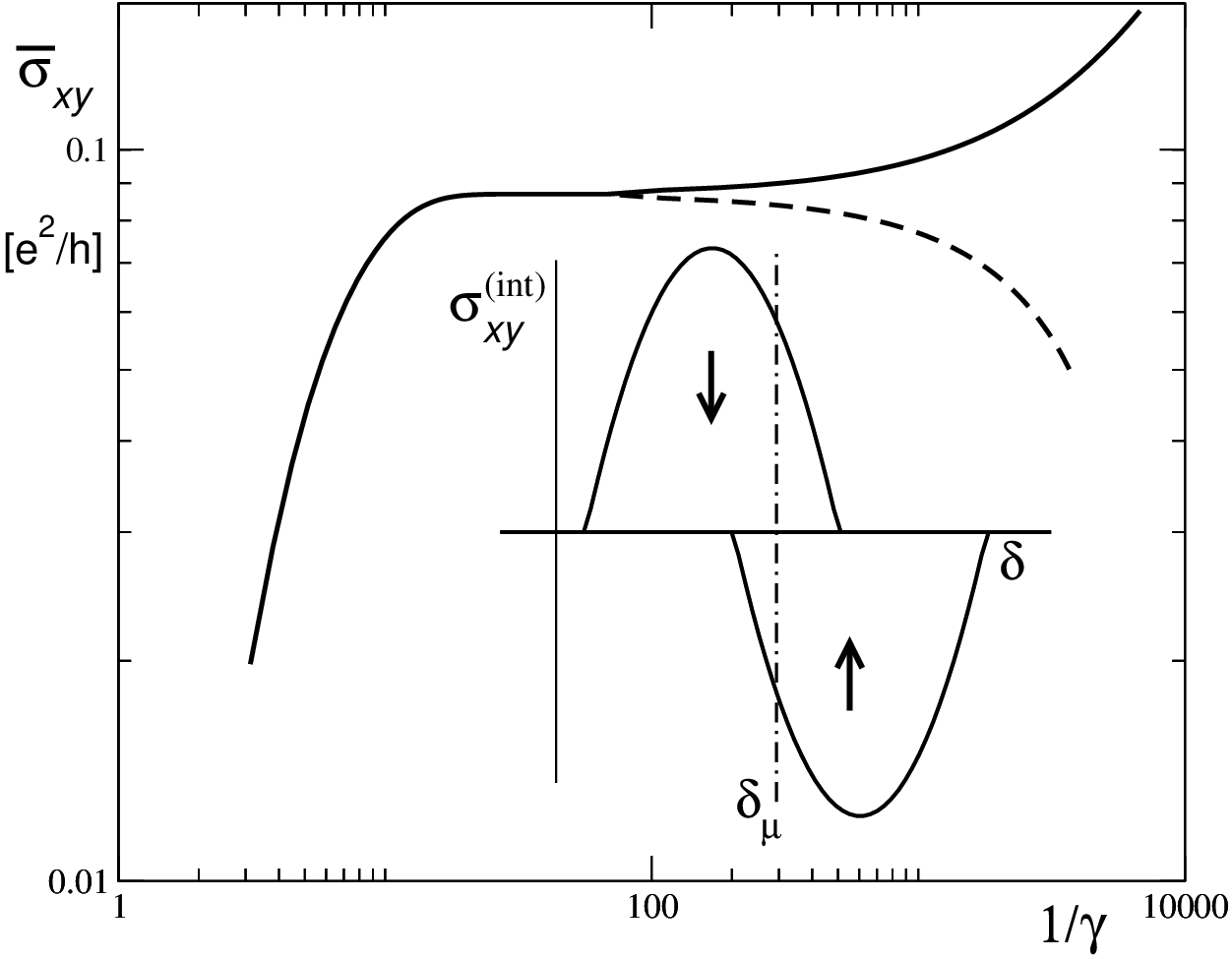}
\caption{Anomalous Hall conductivity for two band model as
function of the disorder strength represented by the parameter
$\gamma$ ($|t|^2=0.3$). Recall that $\sigma_{xx}\propto 1/\gamma$
and width of the unperturbed band is taken
as its unit. Inset shows $\delta$-dependence of the intrinsic
$\sigma_{xy}^{\rm (int)}$ of both bands.}
\label{ahc2b}
\end{figure}

To illustrate qualitative features of the measured anomalous Hall conductivity
dependences on disorder strength covering all three regions the overlap
of energy bands has to be taken into account. For the considered two
dimensional network model two bands having opposite orbital momentum
as well as spin orientation will only be considered for simplicity.
Corresponding intrinsic Hall conductivities have opposite sign but
their absolute values are supposed to be the same as shown in the
inset of the Fig.~\ref{ahc2b}. Their shift due to exchange
interaction is approximated by a Zeeman splitting to obtain a
nonzero Hall conductivity given by the sum of both band contributions
\begin{equation}
\bar{\sigma}_{xy}(\delta_{\mu}) = \left\langle \sigma_{xy}^{\downarrow}
(\delta_{\mu}) \right\rangle_{\rm av} + \left\langle
\sigma_{xy}^{\uparrow}(\delta_{\mu}) \right\rangle_{\rm av}
\; .
\end{equation}
Effect of the disorder will be approximated by a potential energy
fluctuations. Assuming their Gauss distribution the ensemble
averaging reads 
\begin{equation}
\left\langle \sigma_{xy}^{\downarrow,\uparrow}(\delta_{\mu})
\right\rangle_{\rm av} = \frac{1}{\gamma \sqrt{2 \pi}} \int
e^{\frac{(\delta-\delta_{\mu})^2}{2 \gamma^2}}
\sigma_{xy}^{\downarrow,\uparrow}(\delta) \,
d \delta
\; ,
\end{equation}
where the dimensionless parameter $\gamma \sim \Gamma = \hbar/\tau_e$
in units of the unperturbed band width, Eq.~(\ref{gamma}), is assumed
to be energy independent. Sum of both Hall conductivities decreases
with increasing band overlap caused by the band broadening and for
considered $|t|^2=0.3$ $\sigma_{xy} \propto \gamma^{-1.75}$ as shown
in Fig.~\ref{ahc2b}. Unperturbed band separation and Fermi level
position are sketched in the inset. Within intrinsic region the
effect of the band broadening vanishes.

Sum of quantum friction contributions of both bands, Eq.~(\ref{fric}),
multiplied by $\gamma$ has been taken as a fitting parameter.
Linear dependence of $\tau_{qf} \propto \lambda_c$ on a relaxation
time $\tau_e \sim 1/\gamma$ representing longitudinal conductivity has
been assumed through the whole range of disorder strength. It has been chosen
to obtain the experimentally observed range of the intrinsic region
covering approximately two orders of $\sigma_{xx} \propto 1/\gamma$ as
presented in Fig.~\ref{ahc2b}. This assumption is too simple to
illustrate effect of quantum friction precisely.
Like in the classical Couette flow the friction desires some time
to evolve. If it is much larger than the relaxation time $\tau_{qf}$
no effect can be expected. For this reason  the transition
between intrinsic and high conductivity regions should be sharper. 

Under conditions for which electrons are skimming along strip
edges the friction contribution enhances Hall conductivity.
They are expected to take place when electrons are orbiting
close to atomic nuclei and are only tight-bounded to their neighbours.
This is typical for $d$-states (considering most transition metals, for
example, the $s$-states do not contribute to the AHE) and
orbitals can only be slightly perturbed by the surface. 

In exceptional cases,\cite{Miyasato}  
the Hall conductivity even changes its sign upon further decrease of
the dissipation. Such behaviour can be explained within our analysis 
for electrons skipping along surfaces; the friction contribution has
opposite sign and Hall conductivity decreases as shown in
Fig.~\ref{ahc2b} by the dashed line. Skipping orbits appear if
electron orbitals within bulk are of the radius larger than the
interatomic distance of if they are orbiting around interstitial
positions, i.e. $|t^2|>0.5$ and this can occur in gapped
materials.\cite{Correa:2018_a} 
Possibility that the coherence length of minority electrons within the
upper band is larger than that within lower band cannot also be excluded
as origin of this effect.

Despite of the model simplicity it gives qualitative features of
scaling relations between anomalous Hall conductivity and longitudinal
one. It is result of the competition between Hall currents of all
overlapping bands. They are spin polarized and in high conductivity
regions one of them dominates because of the quantum friction effect.
Consequently the resulting Hall current becomes strongly spin polarized.

\section{Summary and concluding remarks}

Existence of the quantum friction in fully coherent systems is
the main message of our treatment. The basic condition of its
appearance is chirality of edge current paths within stripe samples.
This effect persists even in not fully coherent systems for
which the resulting Hall current enhancement is determined
by the coherence length $\lambda_{c}$. This extrinsic
contribution to the measured anomalous Hall conductivity
due to the finite sample dimensions dominates in high
conductivity regime. It represents a quantum analogue of
the classical Couette flow in fluids. It can be expected that
it influences the observed spin Hall effect\cite{Omori:2019_a} in
a similar way.

Presented view to the origin of the Hall current enhancement suggests
that all scattering events affect all conductivity components but they
do so with different efficiency. Electrons flowing along the voltage
drop are subjected to dissipative processes which can be characterized
by a relaxation time $\tau_e \propto \lambda_e$ determining longitudinal
conductivity. On the other side the enhanced Hall current decreases
with disrupting the effective wave interference responsible for
the coupling between current paths. Corresponding
relaxation time $\tau_{qf} \propto \lambda_{c}$
can thus be substantially different from $\tau_e$. Temperature dependent
inelastic scattering of electrons by phonons and magnons is destroying
phase coherence but its effect to $\tau_e$ is weaker. This is consistent
with existence of the intrinsic Hall conductivity plateau.
Also an increase of the diffusion scattering of electrons at sample
surfaces gives rise to the much larger suppression of $\tau_{qf}$
than $\tau_e$ as observed on high conductivity Ni samples. \cite{Xu}
Ratio of the electrical and thermal conductivity
components has been studied for pure Fe samples\cite{Shiomi} doped by
Co and Si. In the limiting case of vanishing temperature
where residual resistivity dominates the validity of the Wiedemann-Franz
law has been confirmed for ratios of diagonal as well off-diagonal
components. It indicates that elastic scattering affect all
components in a similar way and the much shorter $\tau_{qf}$ is
proportional to $\tau_e$. On the other side a more complicated
relation between both times can be expected for inelastic scattering.
Unfortunately, there are not enough experimental data for full
understanding of the electron transport in the high conductivity
regions. Further detail investigations of scattering effects in this
regime is thus desired.  A better understanding of the distribution
of the enhanced Hall current across stripe and the possible
dependence of the measured Hall conductivity on the sample width 
can help to map the evolution of the quantum friction
effect in real systems.

Our analysis of quantum friction predicts that the increase
of $\sigma_{AH}$ (anomalous Hall conductivity) in the high-conductivity
regime should be sensitive to the Hall bar width $w$. It is
supported by experimental data for iron samples. Opposite to
thin layers,\cite{Miyasato} an increase of $\sigma_{AH}$ in massive crystals
of comparable conductivity was not observed.\cite{Majumdar:1973_a}
Nevertheless, more detailed measurements of the relationship between
$\sigma_{AH}$ and $w$ are desirable to confirm our prediction and
experimentally exclude skew scattering as the origin of $\sigma_{AH}$
increasing in the high conductivity region. We stress that the
investigation of relationship~(\ref{fric}) calls for measurements on
a set of devices spanning a large range of $w$.

Direction of the Hall current enhancement is controlled by
the orientation of the current
paths just touching strip edges which is determined
by the boundary conditions. At least in cases for which
suppression of the anomalous Hall effect is observed the
analysis of the current distribution at the
sample edges is desirable to verify origin of this effect.
It would be ideal to be able to vary boundary
conditions. If orbitals of magnetic impurities
periodically distributed within non-magnetic
host lattice are of the radius larger than the distance
between atoms this might be possible at least in principle.
In these cases stripe edges can cut orbitals in half
forcing electrons to skip or leave them untouched.
Systems like Bi$_2$Te$_3$ family of topological insulators
with univalent 3d magnetic ions \cite{Chang:2013_a} seem to be
good candidates. In these systems skipping electrons are
giving rise to chiral edge states crossing energy gap responsible
for the observed quantum Hall effect. Creation of such systems with
mixed boundary conditions might lead to new types of spintronic
devices, diodes, allowing current flow along one direction only.
In this case periodic distribution of ions is not necessary
condition.  Although this sounds as science fiction today we believe
that technological progress will allow to realize such systems
in the future.


\acknowledgments
Authors thank V. Drchal and V. \v{S}pi\v{c}ka for valuable comments,
Y. Niimi for advice regarding experimental data,
J. Ku\v{c}era for technical support and head of the department
J. J. Mare\v{s} for ensuring ideal working conditions for partly
retired author P. St\v{r}eda. We acknowledge financial support from 22-21974S
(granted by the Czech Science Foundation).


\begin{appendix}

\section{KUBO FORMULA RESULTS}

Quantum theory of the linear response of unbounded systems to electric
field at zero temperature leads, for diagonal conductivity components,
to the well known Kubo-Greenwood formula \cite{Greenwood}
\begin{equation} \label{sigma_ii}
\sigma_{i i}(\mu) \, = \, \pi e^2 \hbar
\left\langle \, {\rm Tr} \left\{
v_{i} \delta(\mu - H) v_{i} \delta(\mu - H)
\right\} \, \right\rangle_{\rm av}
\; ,
\end{equation}
and for off-diagonal components the following expression
derived by Bastin et al \cite{Bastin}
\begin{eqnarray} \label{sigma_ij}
\sigma_{i j}(\mu) = i \hbar e^2 \times
\qquad \qquad \qquad \qquad \qquad \qquad \qquad \quad
\nonumber \\ \times \! \! \!
\int\limits_{-\infty}^{\mu} \! \!
\left\langle \! {\rm Tr}
\left\{ \delta(\eta - H) \left[
v_{i} \frac{d G^{+}}{d \eta} v_{j} -
v_{j} \frac{d G^{-}}{d \eta} v_{i} \right]
\right\} \! \right\rangle_{\rm av} \! \! d \eta
\, , \;
\end{eqnarray}
where $H$ denotes a single-electron Hamiltonian, $v_i$ are
components of the velocity operator and
delta-function operator is defined as
\begin{eqnarray} \label{G_pm}
\delta(\eta -H) &=& - \lim_{\epsilon \rightarrow 0^+} \,
\frac{G^+(\eta) -G^-(\eta)}{2 \pi i}
\; \; ,
\nonumber \\
G^{\pm}(\eta) &=& \frac{1}{\eta - H \pm i \epsilon}
\, .
\end{eqnarray}
For crystals with substitutional impurities the ensemble
averaging $\langle \cdots \rangle_{\rm av}$ represents averaging
over impurity configuration.
Generally it is a complicated problem \cite{Levin} which
can be simplified by neglecting vertex corrections allowing
to replace averaged product of resolvents $G(z)$ by product
of their averaged operators
\begin{equation}
\langle G(z) \rangle_{\rm av} \equiv \frac{1}{z-H_{\rm eff}(z)}
\; ,
\end{equation}
where $z$ is the complex energy variable. It has the full
crystal symmetry independently on the character of the
scattering events, asymmetric scattering is not an exception.
Effective Hamiltonian $H_{\rm eff}(z)$ is non-Hermitian and
energy dependent but it is analytic in both complex half-planes,
$H_{\rm eff}(z^{\ast}) \, = \, H_{\rm eff}^{+}(z)$.
Its standard form reads
\begin{equation} \label{self-energy}
H_{\rm eff}(z) = H_0 + \Sigma(z) \quad , \quad
\Sigma(z) \, = \, \Delta(z) - i \Gamma(z)
\; ,
\end{equation}
where $H_0=\langle H \rangle_{\rm av}$ represents virtual crystal
and $\Sigma(z)$ is the energy dependent self-energy
determined by the coherent potential approach \cite{CPA},
as the best known theory to estimate effect of alloying.

Inverse value of its imaginary part represents a
finite electron life-time $\tau$. Note, that matrix elements of
$\Sigma(z)$ are diagonal in representation given by eigenfunctions
of the Hamiltonian $H_0$. Using this representation and neglecting
$\Gamma(\eta)$ entering one of the $\delta$-operators in
Eq.~(\ref{sigma_ii}) we get
\begin{equation} \label{sigma_ii_approx}
\sigma_{ii}(\mu) = e^2 \hbar \sum_{n,\vec{k}}
\frac{\left| \langle n, \vec{k} | v_i | n,\vec{k} \rangle \right|^2}
{\Gamma_{n, \vec{k}}(\mu)} \,
\delta(E'_{n,\vec{k}}(\mu)-\mu)
\; ,
\end{equation}
where $E'_{n,\vec{k}}(\eta) = E^{\rm (0)}_{n,\vec{k}} +
\Delta_{n,\vec{k}}(\eta)$, $n$ and $\vec{k}$ denotes band number
and wave vector, respectively. This expression coincides with the
solution of the Boltzmann equation for longitudinal conductivity.

Neglecting vertex corrections in Eq.~(\ref{sigma_ij}) for the Hall
conductivity, using equality $d G(\eta)/d \eta = - G^2(\eta)$
and having in mind that velocity matrix elements are diagonal in
$\vec{k}$ we get
\begin{eqnarray} \label{sigma_ij_approx}
\sigma_{ij}(\mu) =
\frac{e^2 \hbar}{\pi} \sum_{n,n'}^{n \ne n'} \sum_{\vec{k}}
\int\limits_{-\infty}^{\mu} \frac{\Gamma_{n,\vec{k}}(\eta)}
{[\eta-E'_{n,\vec{k}}(\eta)]^2+\Gamma^2_{n,\vec{k}}(\eta)} \times
\nonumber \\
\times 2 \, {\rm Im} \, \left\{
\frac{\langle n, \vec{k} |v_i| n', \vec{k} \rangle
\langle n', \vec{k} |v_j| n, \vec{k} \rangle}
{\left [\eta-E'_{n',\vec{k}}(\eta) + i \Gamma_{n',\vec{k}}(\eta) \right]^2}
\right\} \, d \eta
\; . \; \; \; \; \;
\end{eqnarray}
With decreasing impurity concentration $\Gamma$ decreases as well and
the dominant contributions are those for which $\eta$-values are close
to $E'_{n,\vec{k}}(\eta)$. If there is no band overlap the
energy difference
$\eta - E'_{n',\vec{k}}(\eta) \approx E'_{n,\vec{k}}(\eta) -
E'_{n',\vec{k}}(\eta)$ dominates the denominator value and
$\Gamma_{n',\vec{k}}(\eta)$ can be neglected if it is much
smaller than the energy difference. This approach thus excludes
significant effect of the decreasing impurity concentration
to the Hall conductivity. This conclusion is general since in
the pure crystal limit vertex corrections are vanishing in
principle. Note that in this limit
Eq.~(\ref{sigma_ij_approx}) gives finite values even in the case
of the band overlap.\cite{Yao:2004_a,Chen:2014_a,Manna:2018_a}

Evaluation of the anomalous Hall conductivity for the considered
ideal network model ($\Gamma \rightarrow 0$) is straightforward
since the energy spectrum is for given spin subsystem composed
of non-overlapping bands and we have
\begin{eqnarray} \label{Bloch_Bastin}
\sigma_{xy}(\mu) =
e^2 \hbar \sum_{m,m'}^{m \ne m'} \sum_{\vec{k}}
f_0 \left(E_{m,\vec{k}} - \mu \right) \, \times
\nonumber \\
\times 2 {\rm Im}
\left\{
\frac{\langle m, \vec{k} |v_x| m', \vec{k} \rangle
\langle m', \vec{k} |v_y| m, \vec{k} \rangle}
{\left[ E_{m,\vec{k}}-E_{m',\vec{k}} \right]^2}
\right\}
\; ,
\end{eqnarray}
where $f_0(E-\mu)$ denotes Fermi-Dirac distribution.
Eigenenergies $E_{m,\vec{k}}$ are
functions of the dimensionless $\delta_{\vec{k}}$ defined by
Eq.~(\ref{cr_dispersion}) and velocity operator does not include
spin-orbit term because of the one dimensional character of
electron orbitals, $\vec{v} = -i \hbar \vec{\nabla}_{\vec{r}}/m_0$.
Contributions for $m-m'= \pm 2$ vanish because periodicity of
wave function amplitudes. Dominant contribution originates in
elements with $m-m'= \pm 1$.

\begin{figure}[h]
\vspace{1mm}
\includegraphics[angle=0,width=3.3 in]{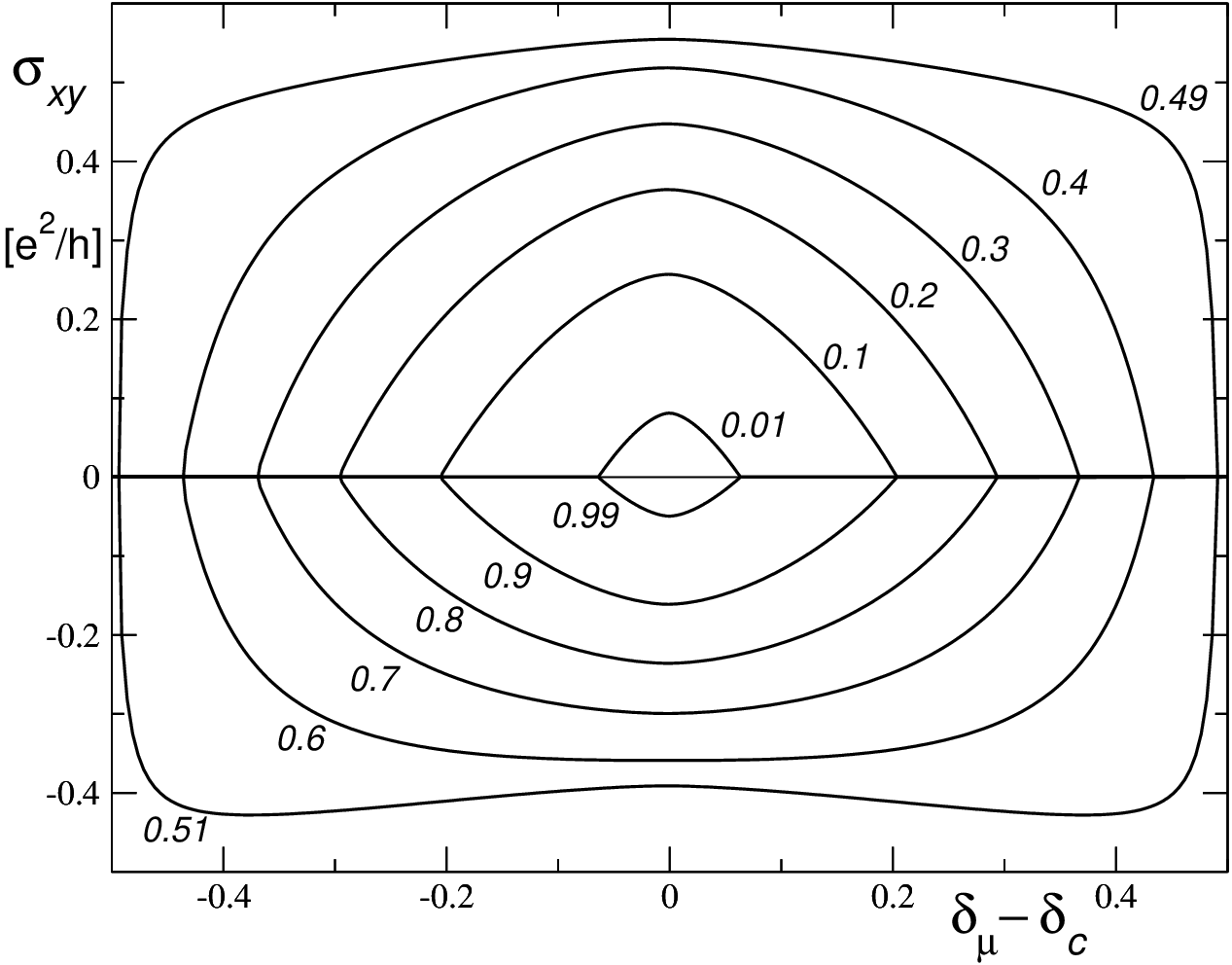}
\caption{Intrinsic anomalous Hall conductivity as function of the
dimensionless
parameter $\delta_{\mu} \sim \sqrt{\mu}$ for several values of $|t|^2$
($\alpha > 0$).}
\label{ahcBas}
\end{figure}

Assuming anti-clockwise motion of electrons on circular orbitals
($\delta > 0$) the obtained anomalous Hall conductivities are shown in
Fig.~\ref{ahcBas} for several values of the transition probability $|t|^2$.
Note that their dependence on dimensionless parameter
$\delta_{\mu} \sim \sqrt{\mu}$
is the same for all bands. Increase of $|t|^2$ above 0.5 changes
sign of the orbital momentum since orbiting of electrons around
interstitial positions becomes dominant. Their average radius is
smaller than that for circular orbitals leading to smaller value of the
orbital momentum. Except of the sign change the lower values of the
Hall conductivity can thus be expected.

For opposite direction of the orbital motion, $\delta \rightarrow - \delta$,
representing subsystem of the opposite spin orientation the
anomalous Hall conductivity changes its sign.
Resulting Hall conductivity is given by the sum of both subsystem
conductivities and its non-zero value can thus only appear if the
spin band degeneracy is removed.

To get longitudinal conductivity the simplest approach reducing
effect of the disorder to an energy dependent imaginary part
$\Gamma(\mu)$ of the self energy will be used
\begin{eqnarray} \label{sig0_Kubo}
\sigma_{0}(\mu) = e^2 \hbar \sum_{\vec{k}} \delta(E_{\vec{k}}-\mu)
\, \frac{\Gamma(\mu) \, | v_x(\vec{k})|^2 }
{(E_{\vec{k}}-\mu)^2 + \Gamma^2(\mu)}
=
\nonumber \\
=
\frac{e^2}{h} \, \frac{1}{2 \pi \gamma(\mu)}
\oint\limits_{F.S.} \left| \frac{d \delta_{\vec{k}}}{d k_x} \right|^2 \,
\frac{d S_{\vec{k}}}
{\sqrt{ \left| \frac{d \delta_{\vec{k}}}{d k_x} \right|^2 +
\left| \frac{d \delta_{\vec{k}}}{d k_y} \right|^2}}
\; ,
\end{eqnarray}
where dimensionless parameter $\gamma(\mu)$ relates to $\Gamma(\mu)$
as follows
\begin{equation} \label{gamma}
\Gamma(\mu) \equiv \frac{4 \hbar^2 \delta_{\mu}}{m_0} \; \gamma(\mu)
\; .
\end{equation}
It has good physical meaning if its value is compared with the band width
represented by the range of available $\delta$ values.


\end{appendix}


\begin{thebibliography}{99}


\bibitem{Nagaosa}
N. Nagaosa, J. Sinova, S. Onoda, A. H. MacDonald, and N. P. Ong,
Rev. Mod. Phys. {\bf 82}, 1539 (2010).

\bibitem{Miyasato}
T. Miyasato, N. Abe, T. Fujii, A. Asamitsu, S. Onoda, Y. Onose,
N. Nagaosa, and Y. Tokura, Phys. Rev. Lett. {\bf 99}, 086602 (2007).

\bibitem{Smit}
J. Smit, Physica (Amsterdam) {\bf 21}, 877 (1955) ; {\bf 24}, 39 (1958).

\bibitem{Luttinger}
J. M. Luttinger, Phys. Rev. {\bf 112}, 739 (1958).

\bibitem{Onose}
Y. Onose, and Y. Tokura, Phys. Rev. {\bf B 73}, 174421 (2006).

\bibitem{Sinitsin}
N. A. Sinitsin, A. H. MacDonald, T. Jungwirth, V. K.
Dugaev, and J. Sinova, Phys. Rev. {\bf B 75}, 045315 (2007).

\bibitem{Onoda}
S. Onoda, N. Sugimoto, and N. Nagaosa,
Phys. Rev. {\bf B 77}, 165103 (2008).

\bibitem{Ishizuka:2017_a} H. Ishizuka and N. Nagaosa,
Phys. Rev. {\bf B 96}, 165202 (2017).

\bibitem{note1}
An explicit expression for this link (between orbital momentum
and AHE) is Eq.~(28) of Wang et al., Phys. Rev. {\bf B 76}, 094406 (2007).
See also review by Yu. Mokrousov (Chapter 6 of
Topology in Magnetism, Springer Series in Solid-State Sciences 192);
time-reversal symmetry can
be broken by spin-orbit, non-collinear magnetic order etc.

\bibitem{Shindou:2001_a} R. Shindou and N. Nagaosa,
Phys. Rev. Lett. 87, 116801 (2001).

\bibitem{Jonckheere}
P. St\v{r}eda, and T. Jonckheere, Phys. Rev. {\bf B 82}, 113303 (2010).

\bibitem{Karplus}
R. Karplus, and J. M. Luttinger, Phys. Rev. {\bf 95}, 1154 (1954).

\bibitem{Adams}
E. N. Adams, and E. I. Blount, J. Phys. Chem. Solids {\bf 10},
286 (1959)

\bibitem{Fivaz}
R. C. Fivas, Phys. Rev. {\bf 183}, 586 (1969).

\bibitem{note2} Term quantum friction is used also by other authors
  in a different context: Volokitin and Persson (Phys. Rev. Lett. 106, 094502)
  discuss van der Waals friction between graphene and an amorphous silicon
  dioxide substrate; Vignale and MacDonald (Phys. Rev. Lett. 76, 2786)
  consider drag in superfluid electron-hole condensates; hydrodynamic
  transport in electron liquids allows to introduce the concept of
  viscosity (Phys. Rev. B 96, 195128). In our case, we focus on
  quantum interference effects in the single-particle picture.
  
\bibitem{Streda_15}
P. St\v{r}eda and J. Ku\v{c}era,
Phys Rev. {\bf B 92}, 235152 (2015).

\bibitem{graphs}
G. Berkolaiko, P. Kuchment: \emph{Introduction to Quantum Graphs},
Amer. Math. Soc., Providence, R.I., 2013.

\bibitem{Landauer57and70}
R. Landauer, IBM J. Res. Dev. {\bf 1}, 223 (1957),
and Phil. Mag. {\bf 21}, 863 (1970).
  
\bibitem{niceREF-QHE}
K. von Klitzing, Rev. Mod. Phys. {\bf 58}, 519 (1986).

\bibitem{Thouless}
D.J. Thouless, M. Kohmoto, M.P. Nightingale,
and M. den Nijs,  Phys. Rev. Lett. 49, 405 (1982).

\bibitem{Exner} P. Exner, J. Phys. A: Math. Gen. {\bf 29}, 87 (1996).

\bibitem{Chalker}
J. T. Chalker and P. D. Coddington,
J. Phys. C {\bf 21}, 2665 (1988).

\bibitem{Marston:1999_a}
J. B. Marston, and Shan-Wen Tsai,
Phys. Rev. Lett. {\bf 82}, 4906 (1999).

\bibitem{Kobayashi}
K. Kobayashi, T. Ohtsuki, H. Obuse, and K. Slevin,
Phys. Rev. {\bf B 82}, 165301 (2010).

\bibitem{Xiao:2010_a}
D. Xiao, M. Chang, and Q. Niu,
Rev. Mod. Phys. {\bf 82}, 1959 (2010).

\bibitem{Berne}
B. A. Bernevig, T. L. Hughes, and S. C. Zhang,
Science {\bf 314}, 1757 (2006).

\bibitem{Konig}
M. K\"{o}nig, S. Wiedmann, Ch. Br\"{u}ne, A. Roth,
H. Buhmann, L. W. Molenkamp, X. L. Qi, and S. C. Zhang,
Science {\bf 318}, 766 (2007).

\bibitem{Chang:2013_a}
Cui-Zu Chang, Jinsong Zhan, Xiao Feng, Jie Shen, Zuocheng Zhan,
Minghua Guo, Kang Li, Yunbo Ou, Pang Wei, Li-Li Wang, Zhong-Qing Ji,
Yang Feng, Shuaihua Ji, Xi Chen, Jinfeng Jia, Xi Dai, Zhong Fang,
Shou-Chen Zhang, Ke He, Yayu Wang, Li Lu, Xu-Cun Ma, and Qi-Kun Xue,
Science {\bf 340}, 167 (2013).

\bibitem{Streda_1994}
P. St\v{r}eda, J. Ku\v{c}era, D. Pfannkuche, R. R. Gerhardts, and
A. H. MacDonald, Phys. Rev. {\bf B 50}, 11955 (1994).

\bibitem{LB} M. B\"uttiker, Y. Imry, R. Landauer, and S. Pinhas,
  Phys. Rev. B {\bf 31}, 6207 (1985).

\bibitem{Kovalev:2008_a} A. A. Kovalev, Karel Vyborny and Jairo Sinova,
  Phys. Rev. B {\bf 78}, 041305 (2008).
  
\bibitem{FP08}
A. Fern\'{a}ndez-Pacheco, J. M. De Teresa, J. Orna, L. Morellon,
P. A. Algarabel, J. A. Pardo, and M. R. Ibarra,
Phys. Rev. {\bf B 77}, 100403 (2008).

\bibitem{Sangiao:2009_a}
S. Sangiao, L. Morellon, G. Simon, J. M. De Teresa, J. A. Pardo,
J. Arbiol, and M. R. Ibarra,
Phys. Rev. {\bf B 79}, 014431 (2009).

\bibitem{Sinova}
Xiong-Jun Liu, Xin Liu, and Jairo Sinova,
Phys. Rev. {\bf B 84}, 165304 (2011).

\bibitem{Streda_2010}
P. St\v{r}eda, Phys. Rev. {\bf B 82}, 045115 (2010).

\bibitem{Naito}
T. Naito, D. S. Hirashima, and H. Kontani,
Phys. Rev. {\bf B 81}, 195111 (2010).


\bibitem{Jungwirth:2002_a} T. Jungwirth, Qian Niu, and A. H. MacDonald,
  Phys. Rev. Lett. {\bf 88}, 207208 (2002).

\bibitem{Yao:2004_a}
Y. Yao, L. Kleinman, A. H. MacDonald, J. Sinova,
T. Jungwirth, Ding-sheng Wang, E. Wang, and Q. Niu,
Phys. Rev. Lett. {\bf 92}, 037204 (2004).

\bibitem{Fuh:2011_a} Huei-Ru Fuh and Guang-Yu Guo,
Phys. Rev. B {\bf 84}, 144427 (2011).

\bibitem{Tung:2012_a} Jen-Chuan Tung, Huei-Ru Fuh, and Guang-Yu Guo,
Phys. Rev. B {\bf 86}, 024435 (2012).

\bibitem{DK13} D. K\"odderitzsch, K. Chadova, J. Min\'ar and H. Ebert,
  New J. Phys. {\bf 15}, 053009 (2013).

\bibitem{Chen:2014_a}
H. Chen, Q. Niu, and A. H. MacDonald,                 
Phys. Rev. Lett. {\bf 112}, 017205 (2014).

\bibitem{Nayak:2016_a} A. K. Nayak, Julia Erika Fischer, Yan Sun,
  Binghai Yan, Julie Karel, A. C. Komarek, Chandra Shekhar, N. Kumar,
  N. Schnelle, J. K\"ubler et al.,
  Sci. Adv. 2, (2016).
  {\tt doi: 10.1126/sciadv.1501870}

\bibitem{Manna:2018_a}
K. Manna, L. Muechler, Ting-Hui Kao, R. Stinshoff, Yang Zhang, J. Gooth,
N. Kumar, G. Kreiner, K. Koepernik, R. Car, J. Kübler, G. H. Fecher,
Chandra Shekhar, Yan Sun, and C. Felser,
Phys. Rev. {\bf X 8}, 041045 (2018).

\bibitem{Helman:2021_a} C. Helman, A. Camjayi, E. Islam, M. Akabori,
  L. Thevenard, C. Gourdon, and M. Tortarolo,
  Phys. Rev. B {\bf 103}, 134408 (2021).

\bibitem{Zhang:2013_a} Y.Q. Zhang, N. Y. Sun, R. Shan, J. W. Zhang,
  S. M. Zhou, Z. Shi, and G. Y. Guo,
  J. Appl. Phys. {\bf 114}, 163714 (2013).

\bibitem{Turek:2012_a} I. Turek, J. Kudrnovsky, and V. Drchal,
  Phys. Rev. B 86, 014405 (2012).


\bibitem{Majumdar:1973_a}
A.K. Majumdar and L. Berger, Phys. Rev. {\bf B 7}, 4203 (1973).
  
\bibitem{Schad:1998_a}
R. Schad, P. Belien, G. Verbanck, V. V. Moshchalkov and Y. Bruynseraede,
J. Phys.: Condens. Matter {\bf 10}, 6643 (1998).

\bibitem{Shiomi}
Y. Shiomi, Y. Onose, and Y. Tokura,
Phys. Rev. {\bf B 79}, 100404 (2009).

\bibitem{Xu} Epitaxial layers of nickel (reported in Phys. Rev. B 85, 220403
  by Xiaofeng Jin and co-workers) can be further improved by adding a copper
  layer on top of the device. While the cap increases the conductivity
  only moderately, larger coherence length due to specular reflection (rather
  than diffuse scattering in the former case) leads to the strong
  increase of $\sigma_{AH}$, see   
  Jianli Xu, Yufan Li, Dazhi Hou, Li Ye, and Xiaofeng Jin,
  Appl. Phys. Lett. {\bf 102}, 162401 (2013).

\bibitem{Correa:2018_a} C.A. Correa and K. Vyborny,
Phys. Rev. B {\bf 97}, 235111 (2018).

\bibitem{Omori:2019_a}
Y. Omori, E. Sagasta, Y. Niimi, M. Gradhand, L. E. Hueso, F. Casanova,
and Y. Otani, Phys. Rev. {\bf B 99}, 014403 (2019).


\bibitem{Greenwood}
D. A. Greenwood, Proc. Phys. Soc. London,
{\bf 71}, 585 (1958).

\bibitem{Bastin}
A. Bastin, C. Lewinner, O. Betbeder-Matibet, and P. Nozieres,
J. Phys. Chem. Solids {\bf 32}, 1811 (1971).

\bibitem{Levin} K. Levin, B. Velick\'{y}, and H. Ehrenreich,
Phys. Rev. {\bf B 2}, 1771 (1970).

\bibitem{CPA}
B. Velick\'{y}, S. Kirkpatrick, and H. Ehrenreich,
Phys. Rev. {\bf 175}, 747 (1968).


\end{thebibliography}
\end{document}